# Acoustooptic Bragg Diffraction in 2-Dimensional Photonic Crystals

*Z.A. Pyatakova*

M.V. Lomonosov Moscow State University, Physics Department

zoya.pyatakova@gmail.com

**Abstract.** The paper shows that silicon-based 2D photonic crystal can be a promising material for acoustooptical devices. Isotropic and anisotropic Bragg diffraction of light in photonic crystal is considered. The computational method for calculation of frequency dependences of Bragg angle is developed. A simple method of optimization of photonic crystal parameters to obtain Bragg diffraction at necessary light and sound frequency is suggested.

**Introduction.** Photonic crystals are promising materials for optoelectronics and optical information processing [1]. Acoustooptics is one of the most efficient ways to control light propagation, and nowadays acoustoopric interaction is implemented in numerous devices. The aim of this work is to estimate the potential of photonic crystals as a material for acoustooptic devices. The interest to photonic crystals is fuelled by two main reasons. First, photonic crystals have unusual dispersion properties [2] that can allow new kinds of acoustooptic interaction. Second, the light and sound in photonic crystals have small group velocity [2], and this fact can lead to enhancement of ligth-sound interaction via enlargement of interaction time and length. The main characteristics of acoustooptical Bragg diffraction are the frequency dependences of Bragg angle and diffraction efficiency. In this work we consider the first aspect, the frequency dependences of Bragg angle. The promising materials that can be used in photonic crystal are silicon with silica rods. The motivation of this choise is that silicon is the most popular material for electronics, but in acoustooptics it is not used of being optically isotropic. Photonic crystals from silicon have large artificial anisotropy, and their papameters can be adjusted to demonstrate unusual dispersion properties in near-IR wavelength range (1.5 μm, the wavelength for optical communications)

**Methods of calculation.** Photonic crystal is modeled as an effective medium for light and sound waves. The vector-diagram technique that implies phase-matching conditions was used to obtain the frequency dependences of Bragg angle [3]. The isofrequency curves of photonic crystals have complex shape that can lead to unusual effects in acoustooptic interaction. Fig.1 illustrates the vector diagrams for 2D photonic crystal.

The vector diagram is a graphical representation of phase-matching condition that is derived from the laws of energy and momentum conservation for photon-phonon interaction.

Energy conservation law:

$$\omega_i = \omega_d + \Omega, \qquad (1)$$

Momentum conservation law:

$$\kappa_i + \mathbf{K} = \mathbf{k}_d + \mathbf{G}. \qquad (2)$$

Indices *i* and *d* rely to incident and diffracted waves, respectively. **k** – wavevector of ligth, **K** – wavevector of sound, ω –light frequency, Ω – sound frequency. In our calculations we neglect the change of frequency in diffraction, because sound frequency is much less than light frequency.

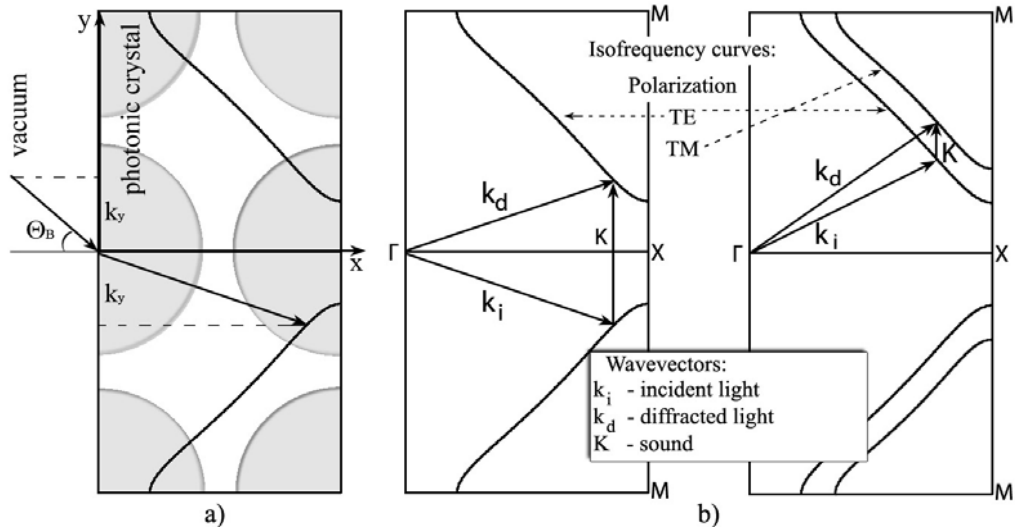

Fig. 1. Vector diagrams for Bragg diffraction in photonic crystals
a) Boundary conditions at the edge of photonic crystal (conservation of tangential component of wavevector). Determination of Bragg angle $\Theta_B$
b) isotropic (without polarization conversion) and anisotropic diffraction (conversion from TE to TM light mode)
Γ, X, M are symmetry points of Brilloin zone of photonic crystal.
In calculations we considered 2D photonic crystal based on Si with cylindrical rods of $SiO_2$, square lattice, filling factor *f* = 0.5.

The dispersion properties for light waves in photonic crystals were calculated by means of plane-wave expansion method (PWE, [4]). To obtain the eigenfrequencies of sound in photonic crystals we used the modified PWE method [5], taking into account the general acoustic anisotropy. The artificial media that have periodic structure with respect to acoustical properties are usually called phononic crystals. Not to mutiply the terms, we use only one term – "photonic crystals". Figure 2 illustrates the dispersion curves for ligth and sound in photonic crystal from silicon with silica rods.

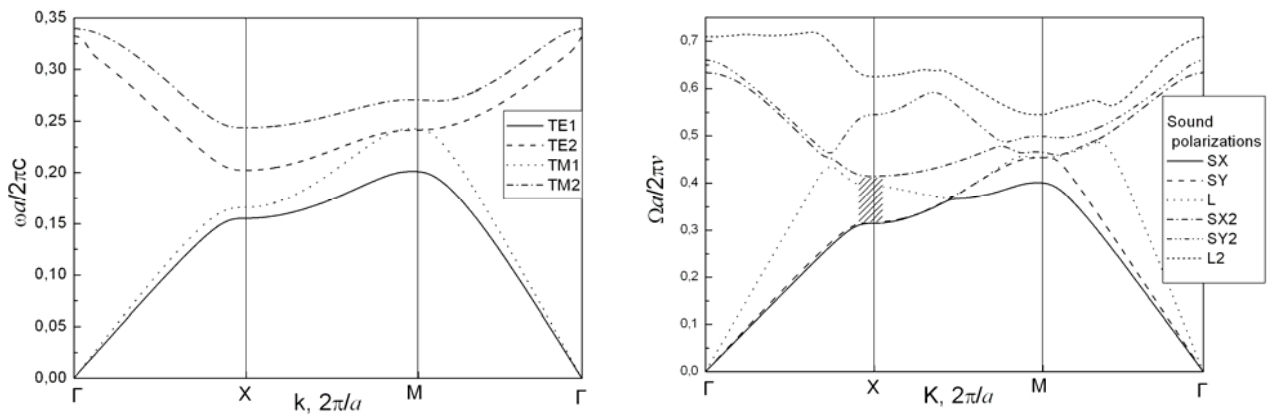

Fig.2. Dispersion curves for light and sound in 2D photonic crystal (Si – $SiO_2$, square lattice, f=0.5) – normalized frequency of light and sound versus normalized wavevector for different polarizations. Light polarizations are TE and TM (TE1,TM1 – 1-st Brilloin zone, TE2,TM2 – 2-nd) and sound polarizations are two quasi-shear (SX, SY) waves and one quasi-longitudinal (L) wave. SX2,SY2,L2 correspond to 2-nd Brilloin zone. Shaded rectangle shows partial forbidden gap for SX waves.

The graphs are shown in normalized coordinates. Normalized frequencies are formulated as:

$$\omega^* = \frac{\omega a}{2\pi c} \text{ for light and } \Omega^* = \frac{\Omega a}{2\pi v} \text{ for sound.} \quad (3)$$

Sound velocity for normalizing was taken $v$=5960 m/s – velocity of longitudinal wave in silica. Normalized wavevector is defined as $k^* = \frac{ka}{2\pi}$. For example, X point corresponds to $k^*=1$.

**Results and discussion**

We consider two kinds of Bragg diffraction – isotropic and anisotropic. Isotropic diffraction in anisotropic material means diffraction without polarization conversion; anisotropic diffraction is followed by polarization conversion. [3]

The dependences of Bragg angle on sound frequency for both kinds of diffraction and different light frequencies are presented on figure 3.

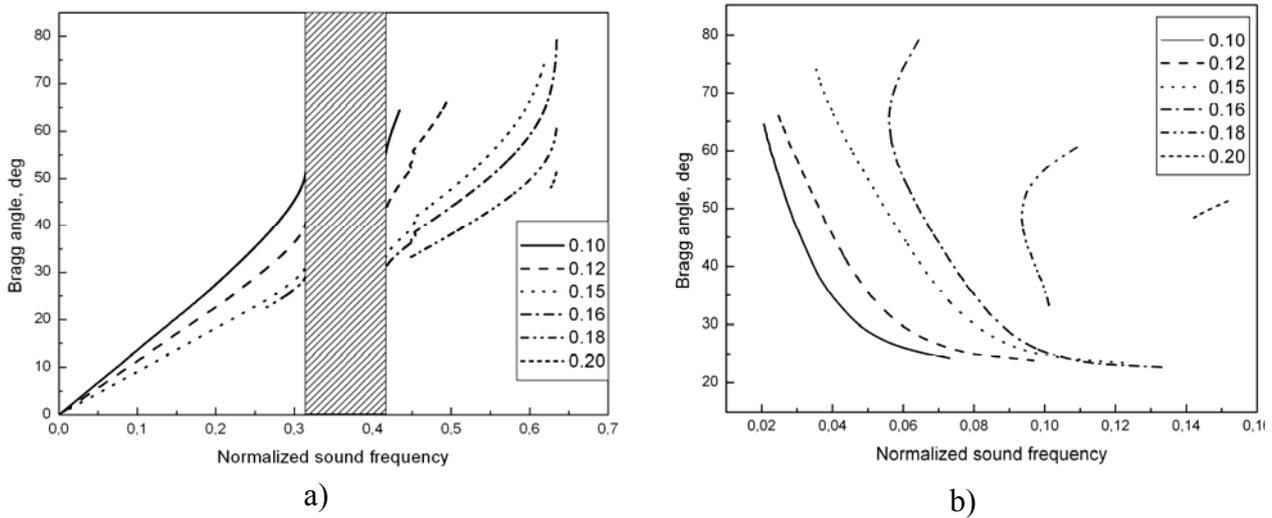

a) b)

Fig.3. Bragg angle frequency dependences for a) isotropic, TE-polarization and b) anisotropic diffraction, TE-TM conversion. Branches correspond to different normalized light frequencies ω*. Shaded rectangle corresponds to the partial forbidden gap (in X-M direction) for acoustic waves of shear polarization.

For the light wavelength much greater than the period of crystal lattice, the frequency dependences of Bragg angle for isotropic diffraction are similar to those of homogeneous crystals. However if light wavelength is close to bandgap, the Bragg angle dependences become ambiguous functions and wide-angle diffraction regimes can exist in a certain frequency range, that can be presicely controlled during the formation of photonic crystal structure. When light wavelength is near the band edge, the range of angles in which diffraction can occur becomes more narrow. This fact can be used for creation filters for image processing.

Now it is time to make a step from normalized parameters to real. For example, we choose the branch of fig.3, b, that corresponds to normalized light frequency ω*=0.16. Suppose operating wavelength of light to be λ=1.5 μm. So, the period of photonic crystal should be equal to $a = \omega^* \cdot \lambda = 0.24$ μm, and radius of silica fiber $r = a\sqrt{\frac{f}{\pi}} = 0.1$ μm.

The minimum sound frequency at which diffraction occurs for the chosen branch is $\Omega_B^* = 0,061$, then $\Omega_B = \Omega_B^* \cdot 2\pi v / a = 9,52 \cdot 10^9$ Hz. This frequency is too high for applications; such frequencies have a strong damping. For isotropic diffraction the frequencies are even higher.

However the parameters of photonic crystal can be adjusted to obtain phase-matching at lower sound frequencies. The way to adjust the parameters is described below.

*Adjustment of photonic crystal parameters*

It is convenient to introduce the effective refractive indices for TE and TM waves by analogy to the refractive index for homogeneous media $n_{TE,TM} = \frac{k_{TE,TM} c}{\omega(k_{TE,TM})}$. In photonic crystal the dispersion law $\omega(k_{TE,TM})$ has to be calculated numerically. Effective refractive index depends on light wavelength and incident angle. With this definition, using the phase-matching condition (2) we can obtain the following expression for sound frequency of anisotropic Bragg diffraction:

$\Omega_B = \omega \frac{v}{c}(n_{TE} - n_{TM})$, where $\Omega_B$ – frequency of sound at which the diffraction occurs, $v$ – velocity of sound, $n_{TE}$, $n_{TM}$ – effective refractive indices for TE and TM waves. In normalized units

$$\frac{\Omega_B^*}{\omega^*} = (n_{TE} - n_{TM}) \quad (4)$$

So the relation of normalized frequencies of light and sound depends only on the difference between effective refractive indices of TE and TM waves for certain Bragg angle.

The question is how to estimate effective refractive indices. First, one should calculate the dispersion law $\omega(k_{TE,TM})$, and in normalized units $n_{eff} = \frac{k^*}{\omega^*}$ (5)

Unfortunately, calculation of dispersion law requires much time and computer resources, and it is not convenient for fast estimation and optimization of parameters. Luckily, it is possible to estimate the effective refractive index from effective medium model. The equations which derive the effective dielectric tensor in matrix media (cylindrical rods, periodically implemented in matrix, square lattice) were obtained in [6]. Here are these equations

$$\varepsilon^{eff}_{11,22} = \varepsilon_1 \left( 1 + \frac{f}{\frac{1-f}{2} + \frac{\varepsilon_1}{\varepsilon_2 - \varepsilon_1}} \right) \quad (6)$$

$$\varepsilon^{eff}{}_{33} = \varepsilon_1 \left(1 + \frac{f(\varepsilon_2 - \varepsilon_1)}{\varepsilon_1}\right)$$

where $\varepsilon_1$ – dielectric permittivity of matrix, $\varepsilon_2$ – of rods, f – filling factor

So, $n_{TE}$ can be estimated as $\sqrt{\varepsilon^{eff}{}_{33}}$, and $n_{TM}$ as $\sqrt{\varepsilon^{eff}{}_{11}}$.

Effective refractive indices depend on refractive indices of constituents and on filling factor. The dependence of $\Delta n = (n_{TE} - n_{TM})$ on filling factor is represented on figure 4.

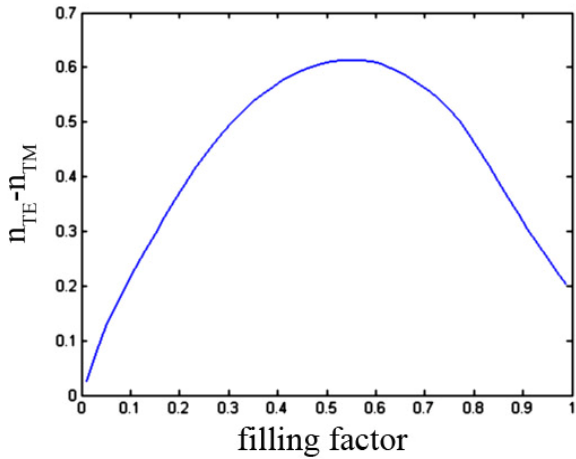

Fig.4. Difference of effective refractive indices vs filling factor for silica rods in silicon matrix.

For example, for the chosen system of silicon with silica rods (*f*=0.5) and wavelength of light λ=1,5 μm the components of dielectric tensor, calculated by means of (6) are $\varepsilon_{11}$=4.62, $\varepsilon_{33}$=7.20. So, *$n_{TE}$ − $n_{TM}$=0.534*. Accurate calculation of effective refractive indices by means of (5) and $\omega(k_{TE,TM})$, presented on fig 2,a for the same system gives *$n_{TE}$ − $n_{TM}$=0.550*. So, the equations (6) allow fast estimation of photonic crystal parameters to obtain anisotropic diffraction at lower sound frequencies.

One can vary the substances of matrix and rods and filling factor to obtain lower sound frequency. For example, assuming filling factor of 0.1 instead of 0.5 (as in our calculations), the difference between refractive indices together with sound frequency is 3 times lower.

*Notes on photoelastic effect.*

Another important characteristic of acoustooptic interaction is diffraction efficiency. But the convensional formulae for diffraction efficiency are not suitable for photonic crystals, because of the large spatial dispersion, that generates problems with definition of dielectric tensor. It is necessary to find another method of calculation. The main effect that defines the diffraction efficiency is photoelastic effect, i.e. change of dielectric tensor under the acoustic wave. Let's consider photoelastic effect in two constituents of photonic crystal under the proper acoustic wave, for example, quasi-shear wave of lowest frequency. Using the PWE method the eigenvector of amplitudes for such wave can be calculated. The classical method of calculation of photoelastic effect in uniform media is shown in [3]

Normalized tensor of elastic deformation is defined as

$$\gamma = \frac{1}{2}\left(\vec{m}\vec{a}^T + \vec{a}\vec{m}^T\right) \qquad (7)$$

where *a* is proper amplitude vector, *m* is wave normal.

The change of inverse tensor of dielectric permittivity can be obtained using the formula

$$\Delta\chi_\mu = A_0 \cdot c_{\mu\nu}\gamma_\nu \qquad (8)$$

where $c_{\mu\nu}$ is 6×6 matrix of photoelastic coefficients.

Assuming the sound intensity to be 10 W/cm$^2$ we obtain

$$\Delta\varepsilon_{Si} = \begin{pmatrix} -0.14 & -0.04 & 0 \\ -0.04 & -0.002 & 0 \\ 0 & 0 & -0.002 \end{pmatrix} \cdot 10^{-4} \text{ for silicon and } \Delta\varepsilon_{SiO_2} = \begin{pmatrix} -0.05 & 0.01 & 0 \\ 0.01 & -0.12 & 0 \\ 0 & 0 & -0.12 \end{pmatrix} \cdot 10^{-5} \text{ for silica}$$

The values of photoelastic effect are comparable to those of conventional acoustooptic crystals (LiNbO$_3$, KDP). It is also seen that non-diagonal components appear in dielectric tensors. So, the TE-TM mixing occurs in photonic crystal.

It is worth noticing that the band structure of photonic crystal is perturbed by photoelastic effect, but this perturbation is substantial only at the edges of band gap. So, for applications in filter systems with light wavelength at the band edge it is necessary to take this effect into account. Also the dispersion properties of photonic crystal change by means of the change of lattice constants, size and shape of rods. So, photoelastic effect in photonic crystal has two mechanisms, first is in changing of dielectric tensor of substances, second is in changing of the geometry of unit cell. Accurate calculation of photoelastic effect is outside in the frames of the present work, but it is, undoubtedly, very important for characterization of photonic crystals as a new material of acoustooptics.

**Conclusion.** Bragg angle dependences of photonic crystal on Si platform show that it is a promising material for acoustooptics. The standard geometry of acoustooptical interaction allows ambiguous Bragg angle dependences that are promising for wide-angle diffraction regime. Considered material allows Bragg diffraction phase matching only at high sound frequencies. High sound frequency is an obstacle for applications, but this problem can be solved by appropriate choice of material parameters, such as filling factor and contrast of refracrive indices.

**Acknowledgements.** Author is thankful to prof. V.I. Balakshy and G.V. Belokopytov for their great help in problem formulation and critical revise of calculation results.